A Robust Deep Learning Method with Uncertainty Estimation for the Pathological Classification of Renal Cell Carcinoma based on CT Images

Ni Yao[1] · Hang Hu[1] · Kaicong Chen[2] · Chen Zhao[3] · Yuan Guo[5] · Boya Li[2] · Jiaofen Nan[1] · Yanting Li[1] · Chuang Han[1] · Fubao Zhu[1] · Weihua Zhou[3,4*] · Li Tian[2*]

[1] School of Computer and Communication Engineering, Zhengzhou University of Light Industry, Zhengzhou 450002, Henan, China

[2] Department of Medical Imaging, State Key Laboratory of Oncology in South China, Guangdong Provincial Clinical Research Center for Cancer, Collaborative Innovation Center for Cancer Medicine, Sun Yat-Sen University Cancer Center, Guangzhou, 510060, Guangdong, China

[3] Department of Applied Computing, Michigan Technological University, Houghton, MI, USA

[4] Center for Biocomputing and Digital Health, Institute of Computing and Cybersystems, and Health Research Institute, Michigan Technological University, Houghton, MI, USA

[5] Department of Radiology, The First People's Hospital of Guangzhou, Guangzhou, 510180, Guangdong, China

*Correspondence:

Li Tian

tianli@sysucc.org.cn

Weihua Zhou

whzhou@mtu.edu

**Abstract**

**Objectives** To develop and validate a deep learning-based diagnostic model incorporating uncertainty estimation so as to facilitate radiologists in the preoperative differentiation of the pathological subtypes of renal cell carcinoma (RCC) based on CT images.

**Methods** Data from 668 consecutive patients, pathologically proven RCC, were retrospectively collected from Center 1. By using five-fold cross-validation, a deep learning model incorporating uncertainty estimation was developed to classify RCC subtypes into clear cell RCC (ccRCC), papillary RCC (pRCC), and chromophobe RCC (chRCC). An external validation set of 78 patients from Center 2 further evaluated the model's performance.

**Results** In the five-fold cross-validation, the model's area under the receiver operating characteristic curve (AUC) for the classification of ccRCC, pRCC, and chRCC was 0.868 (95% CI: 0.826-0.923), 0.846 (95% CI: 0.812-0.886), and 0.839 (95% CI: 0.802-0.88), respectively. In the external validation set, the AUCs were 0.856 (95% CI: 0.838-0.882), 0.787 (95% CI: 0.757-0.818), and 0.793 (95% CI: 0.758-0.831) for ccRCC, pRCC, and chRCC, respectively.

**Conclusions** The developed deep learning model demonstrated robust performance in predicting the pathological subtypes of RCC, while the incorporated uncertainty emphasized the importance of understanding model confidence, which is crucial for assisting clinical decision-making for patients with renal tumors.

**Clinical relevance statement** Our deep learning approach, integrated with uncertainty estimation, offers clinicians a dual advantage: accurate RCC subtype predictions complemented by diagnostic confidence references, promoting informed decision-making for patients with RCC.

**Key Points:**

• Incorporation of uncertainty into deep learning models improves the interpretability and reliability in RCC subtypes classification compared to conventional methods.

• Our experimental results demonstrate that the rate of correct predictions increases as the uncertainty grade of a prediction decreases, confirming its diagnostic performance.

• Radiologists are provided with classification diagnoses and their associated uncertainty, enhancing the transparency and trustworthiness of the AI model in medical imaging applications.

**Abbreviations**

| | |
|---|---|
| AUC | Area under the receiver operating characteristic curve |
| ccRCC | Clear cell renal cell carcinoma |
| chRCC | Chromophobe renal cell carcinoma |
| CI | Confidence interval |
| CMP | Corticomedullary phase |
| CNN | Convolutional neural network |
| CT | Computed tomography |
| NP | Nephrographic phase |
| PCP | Precontrast phase |
| pRCC | Papillary renal cell carcinoma |
| RCC | Renal cell carcinoma |
| ROC | Receiver operating characteristic |
| RoI | Region of interest |
| VoI | Volume of interest |

## Introduction

Renal cell carcinoma (RCC) accounts for approximately 90-95% of renal cancers[1], and is often classified into several histologic subtypes based on their morphological characteristics. The most common is clear cell RCC (ccRCC), accounting for 75% of the cases. This subtype is notably aggressive with a less favorable prognosis. Following ccRCC, the second and third most frequent subtypes are papillary RCC (pRCC) and chromophobe RCC (chRCC), making up 10-15% and 5% of the cases respectively[2]. The RCC subtypes range from indolent to aggressive, each with associated treatment considerations[3]. In addition, the selection of targeted drug therapy and immunotherapy for advanced tumors is also based on the RCC subtypes[4, 5]. Consequently, accurate pathological classification is crucial for patients with RCC in both prognosis and therapeutic strategies[6].

Percutaneous renal biopsy provides diagnostic value in classifying the most common types of RCC[7]. However, the biopsy is invasive, associated with complications, and may be limited by tumor location and the optimal timing for the procedure. Growing evidence suggests that Computed Tomography (CT) plays a crucial role over the course of RCC treatment, from diagnosing and staging the disease to assessing the response to treatment[8]. However, significant overlaps exist in image-level features between renal tumor subtypes, which complicates subtype classification and introduces inter-observer variability[9]. These clinical challenges point to the need for automated systems to reduce misdiagnosis costs and assist radiologists in medical image interpretation[10].

In recent years, deep learning based on convolutional neural network (CNN) has shown potential in enhancing the efficiency of radiologists for complex medical image analysis, especially in the field of RCC[11–13]. Specifically, deep learning has been primarily applied to medical image detection and segmentation[14–17]. It has also been dedicated to purposeful classification tasks, such as distinguishing between benign and malignant tumors and performing differential diagnosis of the histological subtypes of RCC[18, 19]. However, the automation level and accuracy of these approaches need to be improved. In addition, compared to accuracy, sensitivity, specificity, receiver operating characteristic (ROC) curve, and area under ROC curve (AUC), researchers rarely provide uncertainty estimation in their outputs. Uncertainty provides a measure of reliability for the model's predictions, allowing people to determine the model's confidence in its classification decisions more clearly. Therefore, from an evidence-based medicine perspective, our goal is not only to develop an efficient diagnostic model, but also to pay more attention to ensuring that the model can sensitively identify difficult cases and reduce the workload of radiologists.

Against the above background, we aimed to propose a framework for the automatic pathological classification of RCC based on CT images by combining uncertainty estimation and deep learning, including a target detection model specifically designed to identify areas of renal tumors and a classification model to distinguish three major pathological subtypes of RCC.

## Materials and methods

### Study population

This retrospective study was approved by the Ethics Committee of both the Sun Yat-sen University Cancer Center and the First People's Hospital of Guangzhou, with informed consent waived for all patients.

We reviewed medical records from January 2016 to November 2022 at the Sun Yat-sen University Cancer Center (Center 1) and from February 2017 to June 2023 at the First People's Hospital of Guangzhou (Center 2). The patients included had histopathologically confirmed diagnoses of ccRCC, pRCC, or chRCC (Fig. 1). The following criteria applied to data exclusion: (1) incomplete 4-phase CT scanning (unenhanced, corticomedullary, nephrographic, and excretory phases); (2) poor image quality (such as motion artifacts, metal artifacts); and (3) the presence of ambiguous or contradicting clinical data. Finally, 668 consecutive patients from Center 1 (ccRCC n=395, pRCC n=167, chRCC n=106) and 78 from Center 2 (ccRCC n=49, pRCC n=11, chRCC n=17) were included.

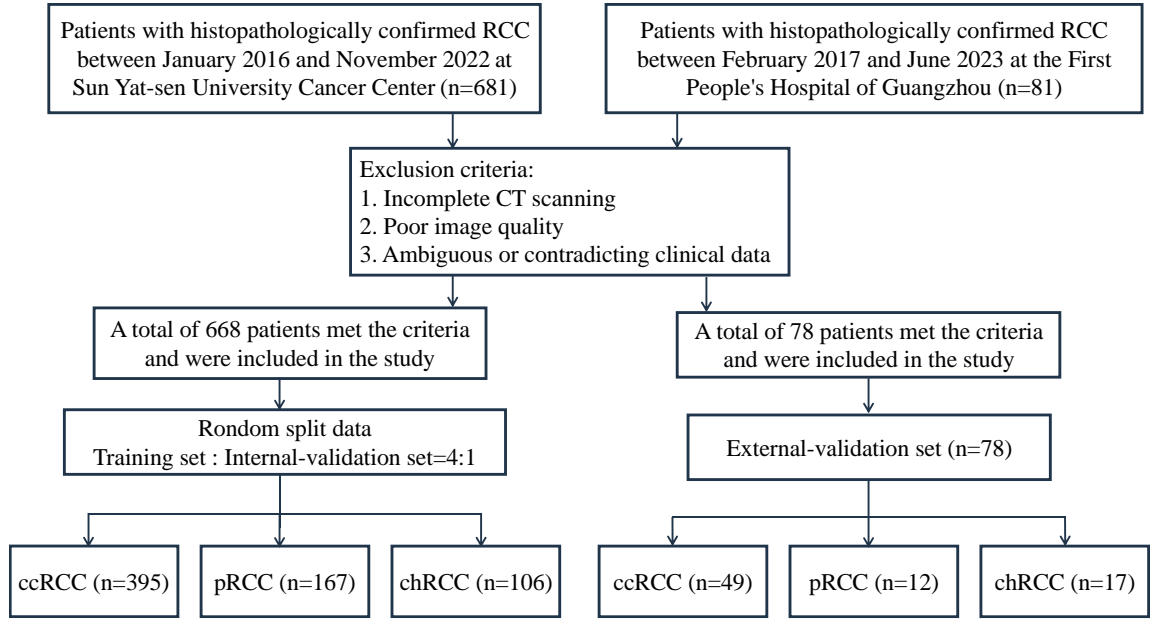

**Fig. 1** Flowchart of patient recruitment. *ccRCC,* clear cell renal cell carcinoma; *pRCC,* papillary renal cell carcinoma; *chRCC,* chromophobe renal cell carcinoma

### Image acquisition

All patients underwent CT examination using multi-slice spiral CT scanners (Center 1: Somatom Force, Siemens Healthineers, Forchheim, Germany; Center 2: Aquilion 64, Toshiba, Tokyo, Japan). Data acquisition was performed in 3 phases: the precontrast phase (PCP), the corticomedullary phase (CMP; 30-second delay after contrast injection), and the nephrographic phase (NP; 90-second delay after contrast injection). After the unenhanced scan, an intravenous bolus injection of nonionic contrast media (350 mg/mL; 1.5 mL/kg) was administered at a rate of 4 mL/s.

### Data preprocessing

The CT slices for each patient were preprocessed to better analyze the renal parenchyma regions, with a window width and level set of 300 and 40, respectively. The CT scans were then resampled to obtain uniform images with a voxel size of $1 \times 1 \times 1\ m^3$. The voxel values were then normalized to the range of 0 to 1.

### Tumor detection

We utilized an automated method to acquire the bounding boxes of the tumors in our dataset. This method allows the model to concentrate on the CT images regions of interest (RoIs) for the subsequent classification of subtypes. A YOLO object detection model using a Pytorch 1.7 framework and Python was built to produce the renal tumor RoIs [20]. This model was trained and validated on the publicly available KiTS19 dataset[21]. Data augmentation techniques including rotation, flipping, and changes in scale to augment the training dataset were used to reduce overfitting. The model achieved an average precision of 0.87 at mean intersection over union (IoU) of 0.5.

CT slices without any kidney tumors were excluded from further analysis. The detection results for each 2D slice were merged to cover the entire tumor in 3D volumes of interest (VoIs). These cropped VoIs were reviewed by a trained technician for validation. They were confirmed by an experienced radiologist and manually corrected if there were mistakes. Then these VoIs were input into our classification model.

**Tumor classification**

The structure of the renal cell carcinoma subtype classification network is shown in Fig. 2. A 3D CNN adapted from the architecture proposed by Wang et al[22] was used. This model incorporates two residual blocks designed for efficient feature extraction and gradient propagation. Each convolutional layer contains multiple channels to extract distinct features from the input, preserving the spatial information from the CT volume. Instead of using a Softmax layer, our architecture outputs non-negative continuous values. These values are treated as evidence, which provides a basis for the parameters of a Dirichlet distribution in our evidence-based uncertainty estimation[23].

The model was trained using a Tesla V100 GPU with 32 GB GPU memory. During the training phase, optimization was performed using an adaptive moment optimizer with a learning rate of 0.0001. The batch size of the models was set to 32, and the number of training epochs was set to 300.

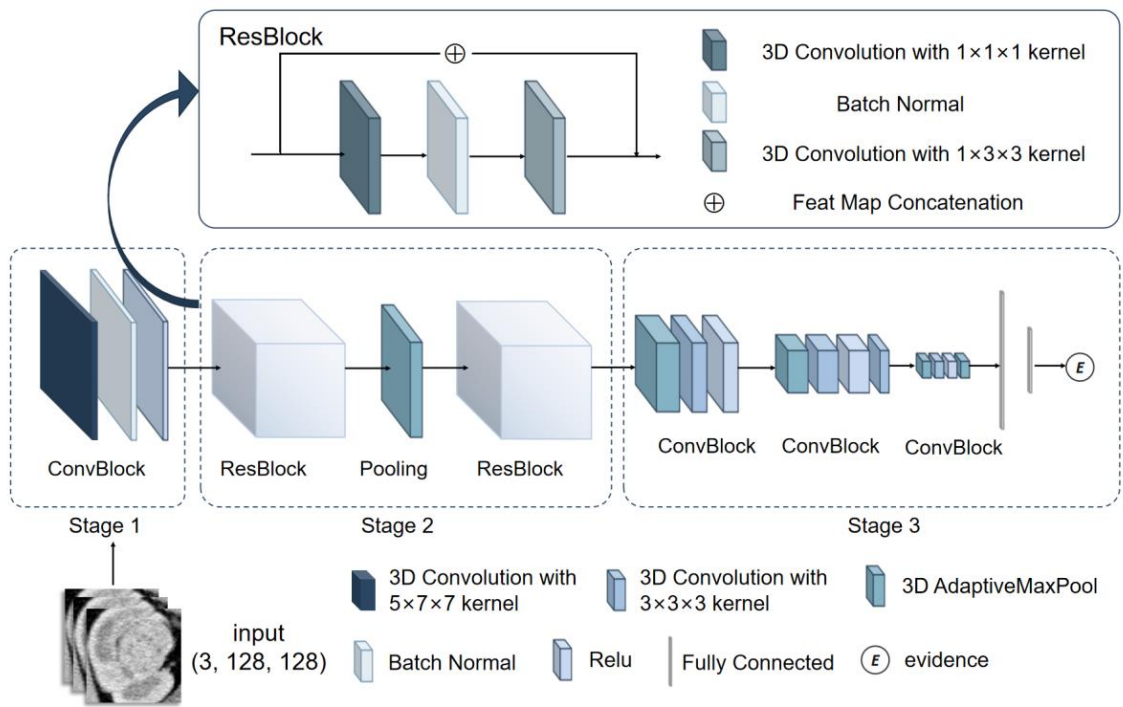

**Fig. 2** Structure of the 3D convolutional neural network for the renal cell carcinoma subtype classification. Stage 1: 3D convolutional block layer; Stage 2: two 3D residual blocks with a pooling layer; Stage 3: output layer producing non-

negative values for uncertainty estimation.

**Uncertainty quantification and loss function**

To quantify the uncertainty in our model predictions, we employed an approach based on the Dirichlet distribution. Central to this method is the translation of our model's outputs, which we term as evidence, into parameters for the Dirichlet distribution. The Dirichlet distribution relies on strictly positive parameters to meaningfully represent the strength or support for different classes. To achieve this, especially when the evidence obtained from our model approached zero, we added a constant of 1 to the evidence. Based on this foundation, the uncertainty $u$ is quantified by the equation:

$$u = \frac{K}{\sum_{j=1}^{K}(e_j + 1)}$$

where $e_j$ corresponds to the evidence associated with the $j$-th class. $K$ representing the total number of classes, serves to normalize the uncertainty measure, ensuring that it remains proportionate and avoids inflate with an increase in the number of classes. In our study, $K$ is 3.

Another crucial component of our uncertainty quantification involves the design of the loss function. The selected loss function serves a dual purpose: it aligns the model's output with the Dirichlet distribution for uncertainty estimation while addressing and mitigating issues related to class imbalances. Class imbalances arise when certain classes have considerably fewer samples than others, potentially causing the model to exhibit a bias toward predicting classes with larger sample sizes. In our datasets, there is a notable imbalance with a greater number of ccRCC samples compared to the other two classes. Our loss function is expressed as:

$$L_i = \sum_{j=1}^{K} y_{ij} \left( \psi\left( \sum_{j=1}^{K}(e_j + 1) \right) - \psi(e_{ij} + 1) \right) w_j$$

Where $i$ represents the $i$-th subject, $y_{ij}$ is the one-hot encoding of the actual class for the $i$-th subject.

The $w_j$ represents the weight for the $j$-th class in the loss function. In this method, it is chosen to be the reciprocal of the number of samples for its respective class. The function $\psi(\cdot)$ is the logarithmic derivative of the gamma function, transforming the model's outputs into probabilistic estimates, thereby facilitating the expression of prediction uncertainty.

**Evaluation**

The evaluation metrics consisted of two parts: the evaluation metrics of RCC subtyping, and the metrics to analyze the effectiveness of introduced uncertainty grades.

The primary classification evaluation metric for both internal and external datasets was the AUC of the ROC curve, assessing accuracy, sensitivity, and specificity. AUC values from five-fold cross-validation were presented with 95% confidence intervals (CI).

Given that the range of uncertainty spans from 0 to 1, we evenly distributed this range to define five uncertainty grades. To evaluate the effectiveness of incorporating uncertainty into the model, the correct rate for each grade was calculated based on the number of accurate predictions relative to the total number of predictions within that specific grade.

## Results

### Population

Table 1 details the demographics and kidney tumor subtype distribution for 746 patients (mean age 53.3 ± 12.5 years) used in our study. The datasets were sourced from Sun Yat-sen University Cancer Center (training and internal validation sets) and the First People's Hospital of Guangzhou (external validation set). In these datasets, the distribution of RCC subtypes remained consistent, with ccRCC being predominant. The overall tumor diameter was 53.8 ± 30.3 mm; more specifically, the diameters for the subtypes were 51.9 ± 28.3 mm for ccRCC, 61.2 ± 36.9 mm for pRCC, and 54.1 ± 30.2 mm for chRCC.

**Table 1** Patient characteristics and tumor size distribution on training and validation sets

| Characteristic | All | Training dataset | Internal validation dataset | External validation dataset |
|---|---|---|---|---|
| Patients, $n$ | 746 | 534 | 134 | 78 |
| Age, $n$ (%) | | | | |
|   <40 | 117 (16) | 92 (21) | 22 (16) | 3 (4) |
|   40-60 | 421 (56) | 295 (68) | 74 (55) | 52 (67) |
|   >60 | 208 (28) | 147 (11) | 38 (28) | 23 (29) |
| Gender, n (%) | | | | |
|   Female | 260 (35) | 181 (34) | 51 (38) | 28 (36) |
|   Male | 486 (65) | 353 (66) | 83 (62) | 50 (64) |
| Subtype, $n$ (%) | | | | |
|   ccRCC | 444 (59) | 316 (59) | 79 (59) | 49 (63) |
|   pRCC | 118 (16) | 85 (16) | 21 (16) | 12 (15) |
|   chRCC | 184 (25) | 133 (25) | 34 (25) | 17 (22) |
| Maximum tumor diameter (mm) | | | | |
|   <30 | 151 (20) | 111 (21) | 26 (19) | 14 (18) |
|   30-50 | 252 (34) | 185 (34) | 39 (29) | 28 (36) |
|   50-70 | 168 (23) | 112 (21) | 34 (25) | 22 (28) |
|   >70 | 175 (23) | 126 (24) | 35 (26) | 14 (18) |

*The training and internal validation datasets are from fold 1 in the five-fold cross-validation procedure. *ccRCC*, clear cell renal cell carcinoma; *pRCC*, papillary renal cell carcinoma; *chRCC*, chromophobe renal cell carcinoma

### Classification performance on validation sets

In the internal validation dataset, the deep learning model achieved AUCs of 0.874±0.039 for ccRCC, 0.849±0.03 for pRCC, and 0.841±0.031 for chRCC. In the external validation set, the AUCs were 0.860 ±0.018, 0.787±0.025, and 0.795±0.029 for ccRCC, pRCC, and chRCC, respectively. Fig. 3 presents the ROC curves for each RCC subtype, based on the model's discrimination for each classification. These curves represent outcomes from the five-fold cross-validation, showcasing the model's performance across different partitions of the dataset.

Additional performance metrics, including accuracy, sensitivity, and specificity for each subtype, are detailed in Table 2. These metrics further provide insight into the model's ability to differentiate

RCC subtypes across the internal and external datasets.

**Table 2** The performance of the model in classifying renal cell carcinoma subtype classification on internal and external validation sets

|  |  | Accuracy | Sensitivity | Specificity | AUC |
|---|---|---|---|---|---|
| Internal Validation Set | ccRCC | 0.791±0.016 | 0.828±0.082 | 0.736±0.111 | 0.874±0.039 |
|  | pRCC | 0.871±0.024 | 0.566±0.042 | 0.929±0.032 | 0.849±0.030 |
|  | chRCC | 0.815±0.030 | 0.634±0.106 | 0.874±0.064 | 0.841±0.031 |
| External Validation Set | ccRCC | 0.763±0.028 | 0.727±0.043 | 0.829±0.064 | 0.860±0.018 |
|  | pRCC | 0.826±0.040 | 0.600±0.050 | 0.864±0.043 | 0.787±0.025 |
|  | chRCC | 0.771±0.027 | 0.600±0.097 | 0.820±0.046 | 0.795±0.029 |

*The data are reported as the mean±SD based on five-fold cross-validation SD standard deviation. *ccRCC,* clear cell renal cell carcinoma; *pRCC,* papillary renal cell carcinoma; *chRCC,* chromophobe renal cell carcinoma; *AUC*, area under the curve.

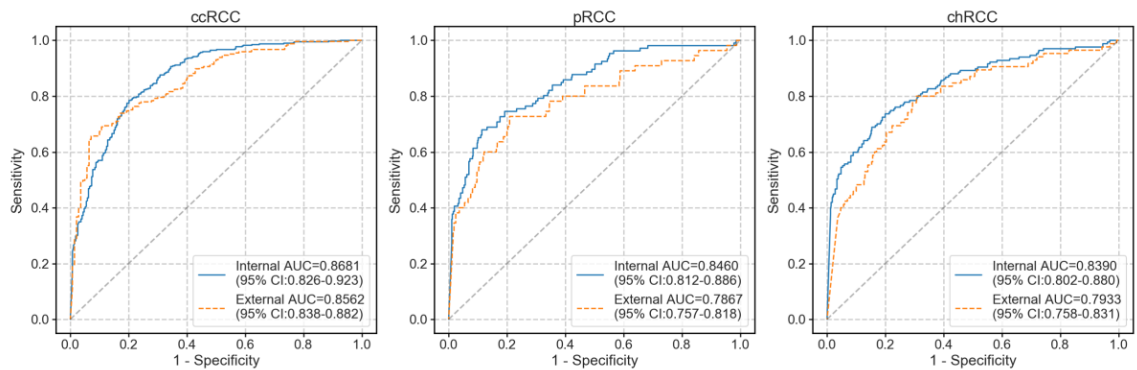

**Fig. 3** The area under the receiver operating characteristic curve (AUC) of five-fold cross-validation in internal and external sets. *ccRCC,* clear cell renal cell carcinoma; *pRCC,* papillary renal cell carcinoma; *chRCC,* chromophobe renal cell carcinoma; *CI*, confidence interval

**Uncertainty analysis**

In our deep learning model, uncertainty estimation was crucial for assessing the reliability of subtype predictions. This evaluation was based on the best model obtained through the five-fold cross-validation.

For the internal validation set, the median uncertainties were 0.24 for ccRCC, 0.22 for pRCC, and 0.19 for chRCC. These values were slightly higher for the external validation set: 0.26 for ccRCC, 0.24 for pRCC, and 0.23 for chRCC. Fig. 4 displays the uncertainty grades and their correlation with correct rates for both validation sets. Overall, as the uncertainty grade decreases, the rate of accurate predictions tends to increase. Fig. 5 contrasts the overall correct rate between the two validation sets across varying uncertainty grades. At grade 1, the correct rate was 89.29% (n=134) in the internal validation set and 77.78% (n=78) in the external validation set. Predictions with high uncertainty grades (4-5, n=5) showed a wrong rate of 80%, which indicates the importance of uncertainty in identifying potentially challenging diagnostic scenarios. Several anomalous cases are illustrated in Fig. 6.

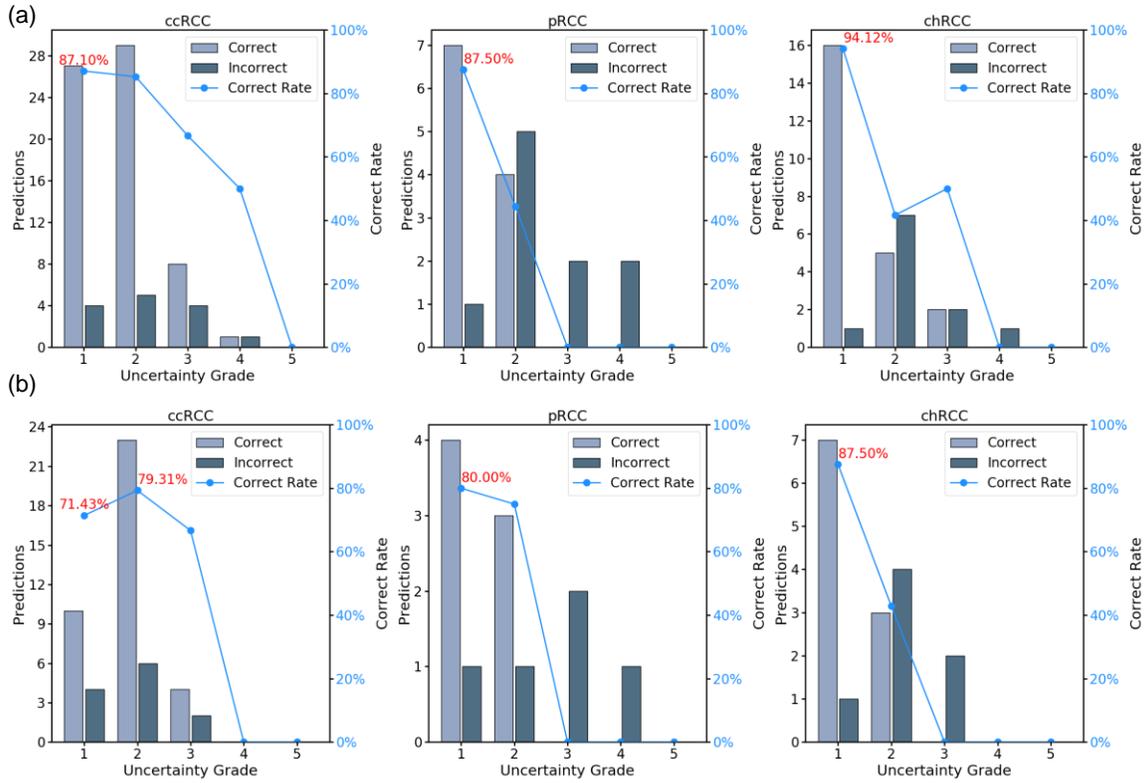

**Fig. 4** Uncertainty-graded prediction performance for kidney cancer subtypes: (a) internal validation set and (b) external validation set. Uncertainty grades 1-5 have an uncertainty range of 0-0.2, 0.2-0.4, 0.4-0.6, 0.6-0.8, and 0.8-1, respectively. *ccRCC,* clear cell renal cell carcinoma; *pRCC,* papillary renal cell carcinoma; *chRCC,* chromophobe renal cell carcinoma.

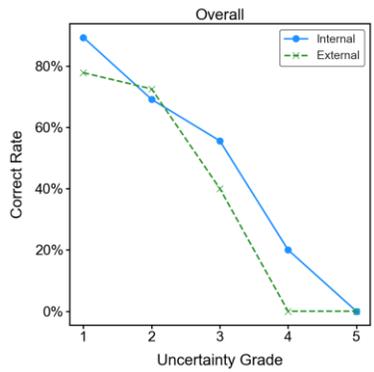

**Fig. 5** Comprehensive correct rate comparison of all subtypes on internal and external validation sets for different grades of uncertainty. Uncertainty grades 1-5 have an uncertainty range of 0-0.2, 0.2-0.4, 0.4-0.6, 0.6-0.8, and 0.8-1, respectively.

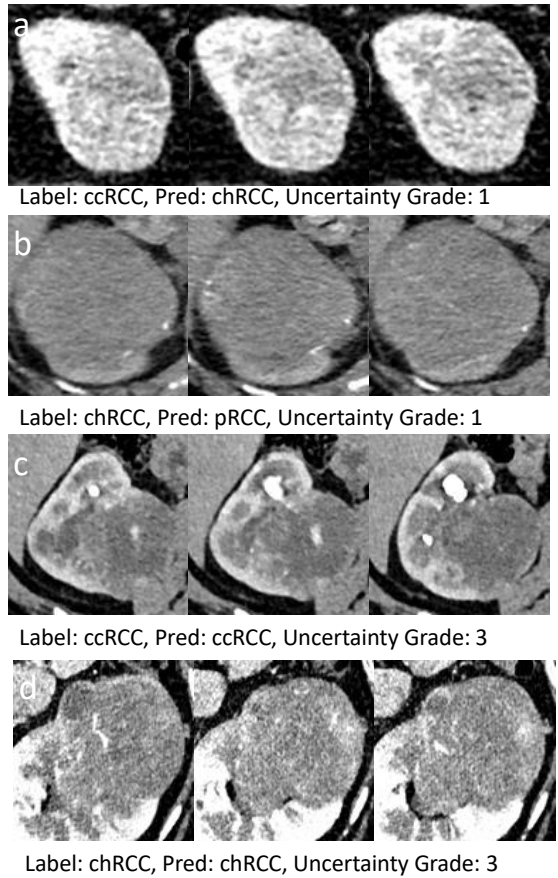

Label: ccRCC, Pred: chRCC, Uncertainty Grade: 1

Label: chRCC, Pred: pRCC, Uncertainty Grade: 1

Label: ccRCC, Pred: ccRCC, Uncertainty Grade: 3

Label: chRCC, Pred: chRCC, Uncertainty Grade: 3

**Fig. 6** Four anomalous cases. The cases have an incorrect prediction with a low uncertainty in a and b; a correct prediction with a high uncertainty in c and d. Uncertainty grades 1-5 have an uncertainty range of 0-0.2, 0.2-0.4, 0.4-0.6, 0.6-0.8, and 0.8-1, respectively. *ccRCC,* clear cell renal cell carcinoma; *pRCC,* papillary renal cell carcinoma; *chRCC,* chromophobe renal cell carcinoma.

**Discussion**

The ability to accurately classify the pathological subtypes of RCC holds profound implications for patient management and therapeutic strategies. In clinical practice, though the accuracy of deep learning-based diagnostic tools is necessary, it is not sufficient. If a model only provides predictions without indicating its uncertainty, it can mislead radiologists especially when diagnosing intractable diseases[24]. This omission can result in radiologists overlooking the need for further tests, potentially impacting patient diagnosis and treatment. In response to this, our study introduces a deep learning-based model with uncertainty estimation for RCC subtype classification. This model not only provides promising accuracy but also incorporates an uncertainty estimation to determine the model's confidence in its predictions.

Incorporating uncertainty in our model distinctly contrasts with traditional models which often provide a sole classification in medical imaging[25–27]. Such an approach can lead to an overreliance on the model, potentially resulting in misdiagnoses, especially when RCC subtypes have similar phenotypic manifestations. Our method not only offers classification but also sheds light on the model's confidence in its prediction. This dual-output offers a new dimension to clinical practice, giving clinicians a more comprehensive understanding of the RCC subtype prediction, thereby aiding in more informed decisions. Moreover, by presenting both the subtype prediction and its associated uncertainty, the model effectively reduces the manual evaluation burden on clinicians. This streamlined process not only makes the diagnostic procedure more efficient but also bolsters the clinician's confidence in adopting the model's recommendations, thereby striking a balance between automation and human expertise.

In our results, a lower uncertainty grade was consistently associated with higher correct rates, a trend was also observed by Song et al.[28] and Ahsan et al.[29]. This demonstrates the advantage of incorporating uncertainty estimation to enhance diagnostic performance. At the lowest uncertainty grade (grade 1), the correct rate for the internal validation set reached 89.29%, and for the external validation set, 77.78%. Thiagarajan et al.[30] employed a low-dimensional visualization techniques and observed that the low uncertainty images revealed distinct class separations, a manifestation of high classification confidence, reinforcing our own observations. However, at the highest uncertainty grade (grade 5), the correct rate for both validation sets dropped to 0. Research by Rączkowska et al suggests that images with the highest grades of uncertainty often display features that are pathologically challenging to classify[31], providing a potential explanation for the observations. Analyzing the data by subtype, the model performed the best for chRCC with correct rates of 94.12% for the internal validation set and 87.50% for the external validation set at grade 1. For pRCC, the rates were 87.5% and 80%, respectively. In contrast, the performance for ccRCC at grade 1 was lower, even though it showed the highest AUC among the three subtypes. One potential factor contributing to this could be the weight settings introduced during model training. Intended to balance the influence of each subtype, the weight adjustments might have unintentionally increased prediction uncertainty for ccRCC. Specifically, the reduced weight for ccRCC, despite its larger dataset, could have made the model more cautious in its predictions for this subtype. This cautiousness might manifest as a higher uncertainty, particularly when the model encounters ccRCC cases that resemble to other subtypes.

Our deep learning model exhibited robust performance across diverse RCC subtypes and validation sets, demonstrating its potential reliability for clinical deployment. While deep learning has been extensively applied in medical imaging, research specifically targeting RCC subtype classification is sparse. Most existing studies primarily focus on distinguishing between benign and malignant tumors, or broader classifications such as differentiating ccRCC from non-ccRCC[32]. Traditionally, RCC classifications have been largely dependent on machine learning methods, which often required significant manual oversight, spanning from data preprocessing to feature extraction[33, 34]. In contrast, our model provides an end-to-end solution, substantially reducing the need for human intervention, enhancing the model's applicability in practice through uncertainty estimation, and serving as a significant step towards the clinical application of automated RCC pathological subtypes diagnosis using deep learning. Although the model in the stage of tumor detection achieved only an average precision of 0.87 at a mean IoU of 0.5, the high IoU is not required since the purpose of this stage is to define a bounding box that covers the tumor. Manual correction can be applied when the bounding box doesn't cover the entire tumor.

Our study has several limitations. First, our study was intended to use CT images from the CMP,

variations in scanning delays occasionally yielded images that leaned toward PCP or NP. The inclusion of these subjects with deviation could potentially influence the model's performance, given that deep learning outcomes are closely tied to the quality and consistency of the training data. Second, while we introduced a measure of reliability by estimating uncertainty, the "black box" nature of deep learning models remains a challenge to interpret. Third, to mitigate data imbalance, we adjusted weights based on the size of each class. While a common approach, this might have unintentionally affected the model's confidence. Last, our research primarily focuses on a specific imaging modality. Future work could benefit from incorporating diverse modalities, such as MRI and histology images, to further enhance our model's diagnostic capabilities.

**Conclusion**

In this study, we developed a deep learning model for pathological classification of RCC, providing valuable support to radiologists. The integration of uncertainty estimates enhances the model's transparency and reliability. Future work will focus on refining the uncertainty estimation mechanism and incorporating multi-modal medical data for improved performance.

**Funding** This study received support from the National Natural Science Foundation of China (Grant Numbers: 62106233, 62303427, and 82370513), and the Henan Science and Technology Development Plan (Grant Number: 232102210010, 232102210062, 232102211003, and 222102210219).

**Declarations**

**Guarantor**    The scientific guarantor of this publication is Li Tian.

**Conflict of Interest**    The authors of this manuscript declare no relationships with any companies whose products or services may be related to the subject matter of the article.

**Statistics and Biometry**    No complex statistical methods were necessary for this paper.

**Informed Consent**    Written informed consent was waived by the institutional review board.

**Ethical Approval**    Institutional review board approval was obtained.

**Study subjects or cohorts overlap**    Study subjects or cohorts have not been previously reported.

**Methodology**
- Retrospective
- Diagnostic or prognostic study
- Multicenter study